# "Actuation at a distance" of microelectromechanical systems using photoelectrowetting: proof-of-concept

Matthieu Gaudet and Steve Arscott[a]

*Institut d'Electronique, de Microélectronique et de Nanotechnologie (IEMN), CNRS UMR8520, The University of Lille, Cité Scientifique, Avenue Poincaré, 59652 Villeneuve d'Ascq, France*

We demonstrate here a proof-of-concept experiment that microelectromechanical systems (MEMS) can be actuated using photoelectrowetting. In order to demonstrate this, a 30 µm thick aluminum cantilever is actuated using an ordinary white light source. A deflection of 56 µm is observed using a light irradiance equal to ≈ 1000 W m$^{-2}$ at a bias of 7 V. The deflection of the cantilever relies on the recently observed photoelectrowetting effect [*Sci. Rep.* **1**, 184 (2011)]. Such "actuation at a distance" could be useful for optical addressing and control of autonomous wireless sensors, MEMS and microsystems.

steve.arscott@iemn.univ-lille1.fr

85.85.+j Micro- and nano-electromechanical systems (MEMS/NEMS) and devices

07.10.Cm Micromechanical devices and systems

68.08.Bc Wetting

68.03.Cd Surface tension and related phenomena

42.79.-e Optical elements, devices, and systems

Microelectromechanical systems (MEMS) now have numerous applications in terms of sensors and actuators.[1] MEMS are commonly actuated[1] piezoelectrically, electrostatically and thermally but also using other techniques such as electrowetting[2-4]. Optical actuation currently relies on radiation pressure[5-8] and photomechanical layers based, for example, on nanowires[9] and shape memory[10]. We demonstrate here a proof-of-concept that MEMS can be actuated using an ordinary white light source via the photoelectrowetting effect[11]. "Actuation at a distance" based on a physical effect could be advantageous for optical addressing and control of autonomous wireless sensors,[12] which currently rely on a circuit-based systems approach,[13] and also autonomous microsystems which are currently under development.[4,14,15]

In contrast to using a photoconductor[16,17] to control electrowetting,[18,19] the photoelectrowetting effect[11] relies on the voltage and optical modulation of the space-charge region of a semiconductor in a liquid/insulator/semiconductor (LIS) junction . Let us consider Fig. 1(a) which shows a droplet resting on a LIS stack. If the semiconductor is grounded, at 0V the droplet will form a contact angle $\theta_0$ with that surface. In the case of p-type semiconductor, placing the droplet to a potential +V will create a depletion region, having a depth $d_s(V,0)$, in the semiconductor. The voltage induced decrease in the capacitance (per unit surface) leads to a modification of the contact angle of the droplet on the surface to $\theta_{V,0}$ [see Fig. 1(b)]. Note that $\theta_{V,0}$ ≠ $\theta_{-V,0}$, this is known as *asymmetrical* electrowetting-on-semiconductors.[11] Irradiation of the semiconductor by photons ($hf > E_{gSi}$) generate electron-hole pairs that modify the depth of the depletion region to $d_s(V,I)$. The effect of this is a decrease of the value of the contact angle to $\theta_{V,I}$, see Fig. 1(c).

The effect can be directly interpreted by a modification of the Young-Lippmann[20,21] equation in the following way:

$$\cos\theta_{V,I} = \cos\theta_0 + \frac{1}{2\gamma}C_{V,I}V^2 \qquad (1)$$

where $\theta_{V,I}$ is the voltage $V$ and light irradiance $I$ dependent contact angle of the liquid, $\theta_0$ is the contact angle at zero bias, $\gamma$ is the surface tension of the liquid. The voltage and light dependent areal capacitance $C_{V,I}$ is given by:

$$C_{V,I} = \frac{\varepsilon_i \varepsilon_s \varepsilon_0}{d_s(V,I)\varepsilon_i + d_i \varepsilon_s} \qquad (2)$$

where $\varepsilon_i$ and $\varepsilon_s$ are the dielectric constants of the insulator and the semiconductor and $d_i$ and $d_s$ are the insulator thickness and the voltage and light dependent thickness of the space-charge layer in the semiconductor respectively. For an LIS junction, both electrowetting and photoelectrowetting will depend upon both semiconductor doping type and density.[11] In classic electrowetting a $\cos\theta \propto V^2$ relationship is observed.[18] In contrast, in a reverse bias LIS junction, $\cos\theta$ is not proportional to $V^2$.[11]

By considering the following equation which is calculated using a geometric approach based on a spherical cap:

$$z = \sqrt[3]{\frac{3V_d}{\pi} \frac{1-\cos\theta}{2+\cos\theta}} \qquad (3)$$

for a constant droplet volume $V_d$, a decrease of the contact angle $\theta$ will be associated with a decrease of the droplet thickness $z$ [see Fig. 1]. In the case of a droplet placed between two surfaces this variation can be used to modify the position of one of the two surfaces.

Let us now consider Fig. 2 which shows a droplet of conducting liquid resting on an insulator/semiconductor stack. The droplet is also in contact with a conducting surface (above) connected to a spring. In equilibrium, the capillary forces, due to the Laplace pressure ($\Delta P = \gamma\kappa$, where $\kappa$ is the curvature of the liquid)[21], will equal the mechanical forces given by Hooke's law ($F_m = -kx$, where $k$ is the stiffness and $x$ is the displacement). We calculate the capillary force to be:

$$F_c = -wl_1\gamma\left(\kappa_1 - \frac{\cos\theta_1 + \cos\theta_{V,I}}{d}\right) \qquad (4)$$

where $\theta_1$ is the contact angle on the conducting surface and $\theta_{V,I}$ the contact angle on the Teflon®/silicon surface depending of the voltage $V$ and the irradiance $I$, $d$ is the distance between the plates, $\gamma$ is the surface tension of the liquid, the liquid wets the conducting surface over a length $l_1$ and $\kappa_1$ is the positive curvature of the liquid surface at the interface liquid/cantilever. The second term in the brackets corresponds to the negative curvature of the liquid calculated with basics trigonometry equations. We know from equation (1) that the contact angle of the droplet can be modified using voltage (asymmetrical electrowetting-on-semiconductors)[11] and light (photoelectrowetting-on-semiconductors[11]). Thus, in principle, we can actuate a mechanical system with light.

If we now consider a cantilever system (see Fig. 3) having dimensions length $l$, width $w$, thickness $t$ and a Young's modulus $E$, we can describe the mechanical restoring force as:

$$F_m = \frac{\delta E t^3 w}{4(l - \frac{1}{2}l_1)} \tag{5}$$

In order to test these ideas we have implemented the simple microcantilever set-up shown in Fig. 3. The aluminum cantilever has dimensions $l$ = 9 mm, $w$ = 2 mm and $t$ = 30 µm and is mounted on a glass support wafer. Taking the Young's modulus of Al to be 70 GPa the spring constant of the cantilever is calculated to be 1.3 N m$^{-1}$. A droplet ($V_d$ = 2.5 µL) of HCL (solution conductivity = 3.64 mS cm$^{-1}$) is placed between the cantilever and the Teflon® surface. The Teflon®/silicon stacks were composed of 20 nm (±3 nm) and 265 nm (± 15 nm) thick Teflon®AF 1600 layers (Dupont, USA) which had been spin-coated (the processing parameters can be found in Ref. 11) onto a commercial single crystal p-type silicon wafer ($N_A$ ~ 1.8×10$^{15}$ cm$^{-3}$) (Siltronix, France). An Al based ohmic contact was formed on the rear surface of the silicon wafer to provide an electrical contact (see Fig. 3). The data was gathered using a commercial Contact Angle Meter (GBX Scientific Instruments, France) (see Supplementary Information) and the experiments were performed in a class ISO 5/7 cleanroom ($T$ = 20°C±0.5°C; $RH$ = 45%±2%). A white light source (KL 2500, Schott, USA) calibrated to 1000 W m$^{-2}$ was used to illuminate the samples. As the silicon wafer is p-type, a positive voltage (+7 V) is applied to the aluminum in order to deplete the silicon at the Teflon®/silicon interface. In this way, the droplet spreads out according to equation 1 with the thickness of the depletion layer $d_s$(V,0) increasing (reverse bias for the LIS structure with the semiconductor p-type).[11] According to equation 4, the variation of the contact angle generates a deflection of the cantilever of a distance $\delta_{V,0}$. Irradiation of the silicon substrate through the insulating layer under the aluminium using a certain angle will generate cantilever deflection $\delta_{V,I}$ in accordance with the model. Fig. 4 shows photographs of the deflection of the Al cantilever/droplet/Teflon®(265 nm)/p-type silicon system as a function of voltage (electrowetting) and light (photoelectrowetting). Fig. 5 shows photographs of the deflection of the Al cantilever/droplet/Teflon®(20 nm)/p-type silicon system as a function of voltage (electrowetting) and light (photoelectrowetting). In both cases the cantilever is successfully deflected using white light. Deflections of 11 µm and 58 µm are observed for Teflon® thicknesses of 265 nm and 20 nm respectively. Table I gives a summary of the experimental results of the study and Table II gives the results from the force modeling using equations 4 and 5. It can be seen that calculations of the mechanical restoring force and the capillary force agree very well. The value of $\kappa_1$ in equation 4 has been fitted to 833 m$^{-1}$, corresponding to a value in accordance with the geometric size of the cantilever.

An analytical model for $d_s(V,I)$ of a MOS capacitor[22] under illumination[23] can be injected into the Young-Lippmann equation (equation 1) in order to calculate the change of the contact angle $\Delta\theta_{V,I}$ due to the light at a given voltage. For the LIS stacks considered here at +7V (Teflon® thickness = 20nm) $\Delta\theta_{7V,L}$ is calculated to be 16.8°; the measured value of $\Delta\theta_{7V,I}$ is 18.1°. At +20V (Teflon® thickness = 265 nm) $\Delta\theta_{20V,L}$ is calculated to be 7.4°; the measured value of $\Delta\theta_{20V,L}$ is 12.5°. In order to perform these calculations one needs to know the transmission of light into the semiconductor for the two Teflon® thicknesses and the subsequent generation rate of electron-hole pairs in the space-charge region. Teflon® is virtually transparent to white light[24] and over the wavelengths 350 nm (violet) to 750 nm (red) the transmission of the light into the semiconductor layer is calculated to be ≈60% for 265 nm and ≈53% for 20 nm.[25] An irradiance of 1000 W m$^{-2}$ over this band equates to a generation rate of >10$^{20}$ cm$^{-3}$ s$^{-1}$ over the entire depletion width of 2.14 µm (20 nm) and 2.52 µm (265 nm) (i.e. the calculated depletion width under dark conditions at +7V and +20V using $\varepsilon_s$ = 11.9, $\varepsilon_i$ = 1.92, $d_i$ = 20 nm and 265 nm).[22] Thus the capacitance of the stack *under illumination* $C_{V,L} \rightarrow C_i$ i.e. the insulator capacitance.[23]

It should be noted that actuation using a liquid/insulator/semiconductor junction ensures low power consumption actuation in a similar way as the MOS junction[22] provides a low power device as actuation is achieved at virtually zero current. Finally, by demonstrating an overlap between MEMS, wetting phenomena and semiconductors we have shown here that silicon-based technologies incorporating these three subjects are feasible.

**Figure Captions:**

**FIG. 1.** Principle of photoelectrowetting-on-semiconductors.

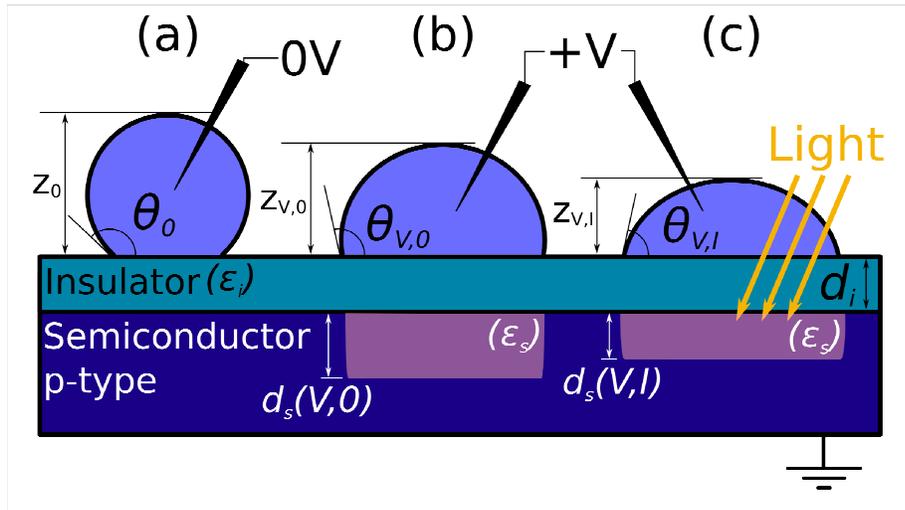

**FIG. 2.** Balance of capillary forces and mechanical restoring force for a droplet between two surfaces.

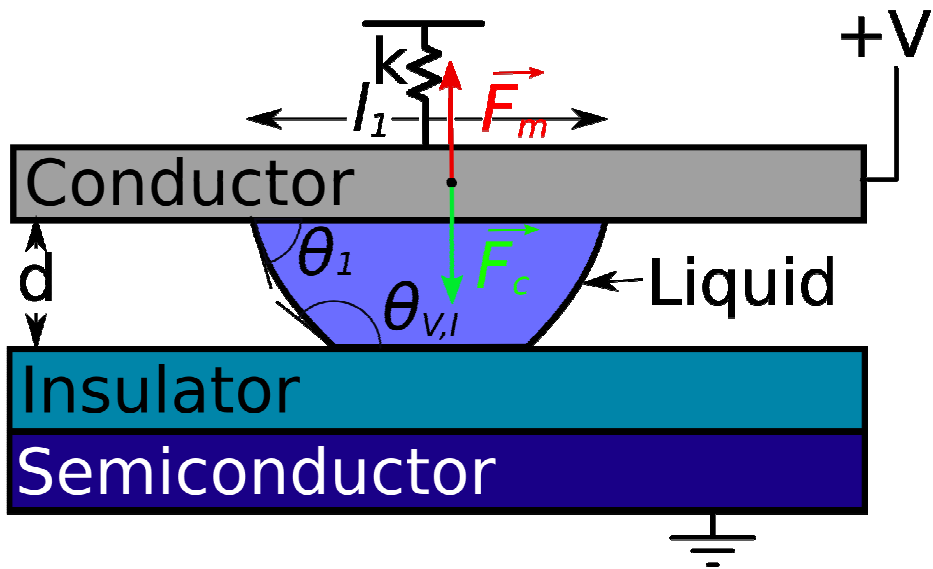

**FIG. 3.** Experimental setup showing aluminum cantilever, liquid droplet ($H_2O/HCl$, $c = 0.01M$) and Teflon® AF/silicon photoelectrowetting surface.

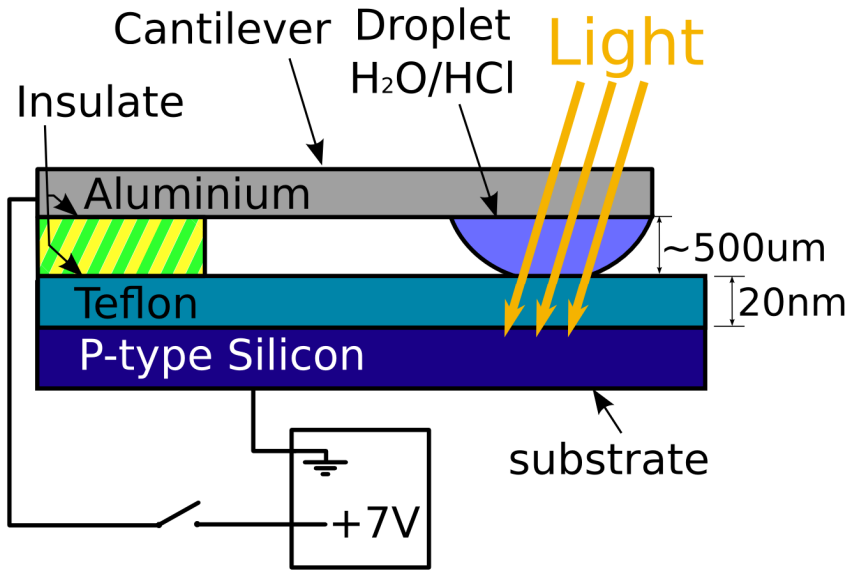

**FIG. 4.** Deflection of a 30µm thick Al cantilever using photoelectrowetting on a Teflon® AF (265 nm)/p-type silicon ($N_A = 1.8 \times 10^{15}$ cm$^{-3}$) stack at (a) 0V (dark), (b) +20V (dark) and (c) +20V (illuminated).

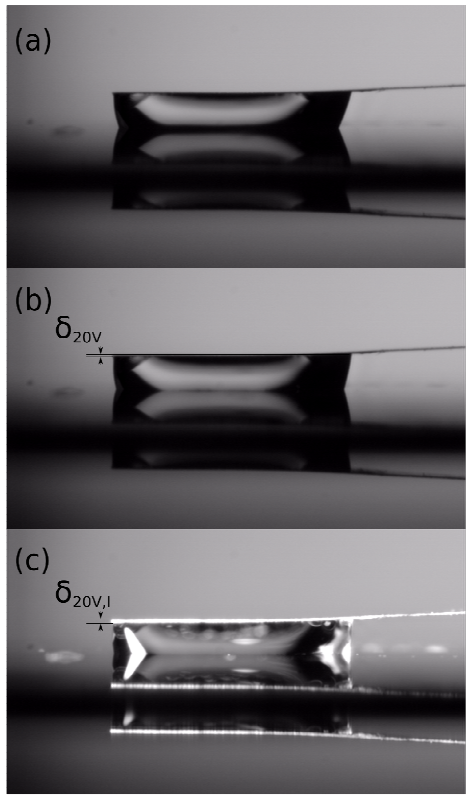

**FIG. 5.** Deflection of a 30μm thick Al cantilever using photoelectrowetting on a Teflon® AF (20 nm)/p-type silicon ($N_A = 1.8 \times 10^{15}$ cm$^{-3}$) at (a) 0V (dark), (b) +7V (dark) and (c) +7V (illuminated).

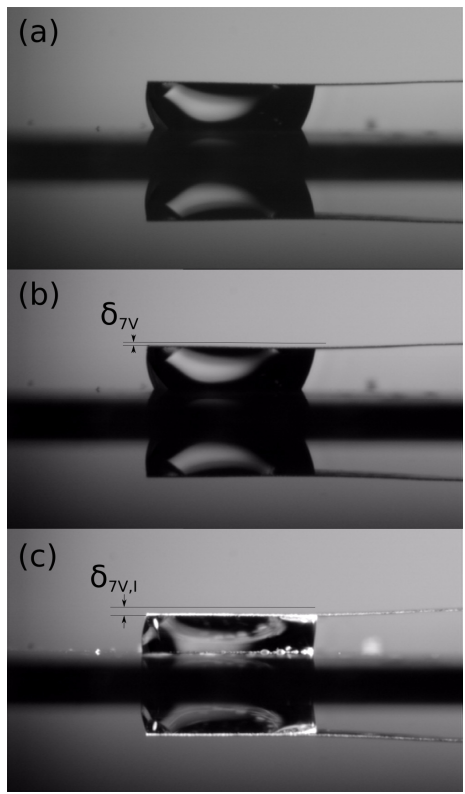

**TABLE I.** Summary of experimental results.

| V | Light | $d$ (μm) | $\delta_{V,I}$ (μm) | $l_1$ (mm) | $\theta_1$ | $\theta_{V,I}$ |
|---|---|---|---|---|---|---|
| 0V | no | 581 | 39 | 2.254 | 82.1° | 119° |
| 7V | no | 542 | 78 | 2.239 | 82.3° | 109.7° |
| 7V | yes | 486 | 134 | 2.278 | 84.8° | 91.6° |
| 0V | no | 440 | 65 | 3.094 | 76° | 116° |
| 20V | no | 409 | 96 | 3.091 | 83.7° | 103° |
| 20V | yes | 398 | 107 | 3.119 | 84.3° | 90.5° |

**TABLE II.** Summary of force balance calculations.

| V | Light | $F_c$ (µN) | $F_m$ (µN) |
|---|---|---|---|
| 0V | no | 77.5 | 75.5 |
| 7V | no | 149.9 | 150.6 |
| 7V | yes | 269.3 | 260.7 |
| 0V | no | 145.9 | 148.4 |
| 20V | no | 220.9 | 219 |
| 20V | yes | 246.5 | 245.5 |